\documentclass[prb,floatfix,aps,twocolumn,showpacs]{revtex4}
\usepackage{graphicx}
\usepackage[]{epsfig}
\usepackage[]{psfig}
\newcommand{\beq}{\begin{equation}}
\newcommand{\eeq}{\end{equation}}
\newcommand{\bea}{\begin{eqnarray}}
\newcommand{\eea}{\end{eqnarray}}
\newcommand{\nn}{\nonumber}

\newcommand{\lam}{\lambda}

\begin{document}

\title{Transport through a quantum dot with SU(4) Kondo entanglement} 
 
\author{K. Le Hur$^{1,2}$, P. Simon$^{3,4}$, D. Loss$^4$}
\affiliation{$^1$ Department of Physics, Yale University, New Haven, CT 06520}
\affiliation{$^2$ D\' epartement de Physique et RQMP, Universit\'e de Sherbrooke, Sherbrooke, Qu\' ebec, Canada, J1K 2R1}
\affiliation{$^3$ Laboratoire de Physique et Mod\' elisation des Milieux Condens\' es, CNRS and Universit\'e Joseph Fourier, B.P. 166, 25 Avenue des Martyrs, 38042 Grenoble, France}
\affiliation{$^4$ Department of Physics and Astronomy, University of
Basel, Klingelbergstrasse 82, CH-4056 Basel, Switzerland}

\date{\today} 

\begin{abstract}
We investigate a mesoscopic setup composed of a small electron droplet (dot) coupled to a 
larger quantum dot (grain) also subject to Coulomb blockade as well as two macroscopic leads used as 
source and drain. An exotic Kondo ground state other than the standard SU(2) Fermi liquid unambiguously emerges: an SU(4) Kondo correlated liquid. The transport properties through the small dot are analyzed for this regime, through boundary conformal field theory, and allow a clear distinction with other regimes such as a two-channel spin state or a two-channel orbital state. 
\end{abstract}

\maketitle

\section{Introduction}
The first experimental observations of the Kondo effect in semiconducting 
quantum dots\cite{Gold,Cron,vdw,kouwenhoven} 
have triggered a huge theoretical and experimental activity in this field during the last years.
It turns out that the Kondo effect is a rather generic feature that may appear 
as soon as a mesoscopic conductor
is strongly coupled to electronic leads. The Kondo effect has thus been observed
when connecting 
molecular conductors such as carbon nanotubes\cite{cn} or large organic molecules\cite{mol} 
to metallic electrodes. These experimental breakthroughs have also brought new questions
concerning the robustness of the mesoscopic Kondo effect like what happens to 
the quantum dot or molecular conductor
when  connected between between ferromagnetic leads\cite{ferroexp} or superconducting 
electrodes.\cite{scexp}

\begin{figure}[ht]
\centerline{\epsfig{file=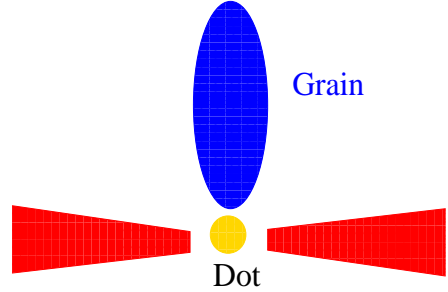,angle=0.0,height=3.8cm,width=5.8cm}}
\caption{(color online) A small dot in the Kondo regime which acts as an S=1/2 spin impurity is weakly coupled to a large dot (grain) as well as to two leads serving as source and drain.}
\label{setup}
\end{figure}

Another exciting feature predicted by earlier theoretical works on the Kondo effect 
is the possibility to reach exotic non Fermi liquid fixed points.\cite{nozieres} 
The most generic one is the so called overscreened fixed point which appears when a spin $S$
impurity coupled to $n>2S$ channels. These exotic fixed points are reached at low energy.
At fixed temperature, or magnetic field or other external parameters, the vicinity
of these fixed points is embodied by non trivial scaling relations involving power laws. These scaling relations may be obtained using the powerful conformal field theory techniques.\cite{AL}
Observing and controlling non Fermi liquid features around a mesoscopic conductor 
constitutes an experimental challenge since these non Fermi liquid fixed point are on one hand 
rather unstable toward various perturbations like channel asymmetry and on the other hand
the associated Kondo scale may be rather small compared to the aforementioned Kondo effect
observed by various groups. Concerning the explicit experimental realization of those issues, a rather simple setup has been envisioned by Oreg and Goldhaber-Gordon \cite{oreg} which is sketched in Fig. \ref{setup}.

The device consists of a single quantum dot which can be tuned in the Kondo regime.
This small quantum dot is coupled by a tunnel junction to a larger quantum dot.
The most important feature is that this larger quantum dot (dubbed as the grain hereon)
has some Coulomb energy $E_c$. The number of electrons inside this grain is 
controlled by a plunger gate voltage $V_g$. Although this larger quantum dot is
of micron metric scale, it can still behave  as a good single-electron transistor. The small quantum dot is weakly-coupled to bulk leads, which allows to perform transport measurements across the small dot while the grain charge remains to a large extent quantized.

When the grain is in a Coulomb blockade valley, direct tunneling electronic processes between the leads and the grain are not allowed for temperatures below the Coulomb energy $E_c$.
Therefore the grain acts as an independent channel and a linear combination of the two leads provides
another distinct channel. If the dot is tuned in the Kondo regime, such a geometry would therefore
be a good candidate to reach a $2$-channel Kondo fixed point provided the temperature is low enough 
and that the grain level spacing is much smaller that the 2-channel Kondo temperature.
\begin{figure}[ht]
\centerline{\epsfig{file=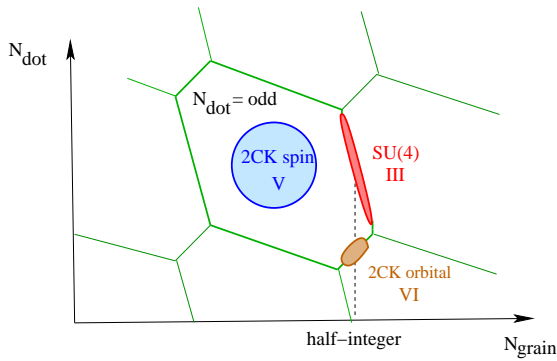,angle=0.0,height=4.7cm,width=7.3cm}}
\caption{(color online) Stability diagram of the dot-grain system. The colored areas indicate where the the novel Kondo fixed points may be found with some fine-tuning of the tunneling amplitudes. Three fixed points are distinguished: 2CK spin (2-channel Kondo spin), 2CK orbital (2-channel Kondo orbital), and SU(4) Kondo are discussed in the sequel. Here, III, V, VI refer to the sections in the text, the mean number of electrons on the dot (grain) is denoted $N_{dot}$ $(N_{grain})$, and each domain is embodied by a different set $(N_{dot},N_{grain})$.} 
\label{phasediagram}
\end{figure}
It has been theoretically predicted that the conductance through the small dot reach $e^2/h$ 
for symmetrically coupled leads and that at zero temperature deviations 
should follow a square root  behavior in the bias voltage (which is quite different from a Fermi liquid fixed point from deviations scale quadratically).\cite{Pustilnik} It turns out that some square root deviations have been indeed observed as a function of the bias voltage.\cite{David} 
 
Although the setup depicted in Fig. \ref{setup} is rather simple, it exhibits a very rich phase 
diagram, as summarized in Fig. \ref{phasediagram}, and many other exotic regimes can {\it a priori} be reached especially when the grain is tuned to resonance between two charge states (the two charge states become degenerate) resulting in half-integer values of $N_{grain}$.

Already when a single grain is connected
directly to a lead, it has been predicted by Matveev \cite{Matv1}
that the charge fluctuations in the grain at the degeneracy points behave as a pseudo-spin.
Indeed, Matveev mapped this problem  of charge fluctuations 
onto a 
(planar)
two-channel Kondo Hamiltonian with the two charge 
configurations in
the box playing the role of the impurity spin\cite{Matv1,Georg} and the 
physical spin of the
conduction electrons acting as a passive channel index. 
Unfortunately, these predictions have never been observed in a convincing manner, even though
some concrete endeavors in this direction have been performed,\cite{Konrad}
essentially because the Kondo temperature of this 2-channel charge Kondo effect has to be 
much larger than the grain level spacing which is quite a stringent constraint.
It turns out that the device depicted in Fig. \ref{setup} may offer a better chance
for observing these two-channel charge Kondo behavior. It has indeed been shown that  
when the tunneling between the lead and the grain involves a 
small dot at resonance it may actually considerably enhance 
the Kondo temperature of the system.\cite{Schiller} Even a line of $2-$channel fixed points
mixing spin and charge degrees of freedom has been numerically predicted for arbitrary dot Coulomb
interaction.\cite{Anders} 

Another interesting possibility occurs when the small dot is tuned into a Kondo regime whereas 
the grain is tuned to a charge degeneracy point. It has been shown that the low energy 
physics can be described by an SU(4) Fermi liquid fixed point where the spin degrees of freedom of the small dot are entangled with the charge fluctuations of the grain.\cite{Karyn}
Contrary to the aforementioned fixed point, this SU(4) fixed point is marginally stable.
Furthermore, it is embodied by an enhanced Kondo temperature compared for example to the
usual SU(2) Kondo effect. Let us mention that the possibility of a strongly correlated 
Kondo ground state possessing an SU(4) symmetry has also been discussed very
recently in  double dots coupled with a strong capacitive
inter-dot coupling where orbital and spin degrees of freedom are
intertwined\cite{Borda,Pascal_Rosa_Karyn,Galpin,Eto} as well as in triangular dots.\cite{Brataas} An
SU(4) Kondo effect seems to have been observed in carbon nanotubes where the spin degrees of freedom are entangled with both chiralities of the carbon nanotubes.\cite{KG}
Nevertheless these quantum numbers are locally entangled inside the whole nanotube geometry.
By analyzing the transport properties through this whole geometry, it is
extremely difficult to differentiate between an ordinary two-level SU(2) Kondo effect from the SU(4) Kondo state when based only on linear transport.\cite{Rosa1,Rosa2}
In this sense, in the geometry depicted in Fig. \ref{setup}, the spin degrees of freedom
being spatially separated from the grain charge excitations allows a more direct probe
as stressed below.

Our paper is structured as follows: In Section II, we present our modeling of the setup
depicted in Fig. \ref{setup}. In Section III, we summarize the main steps
leading to this SU(4) kondo effect for completeness. Section IV is the core of this paper
and is devoted to transport properties through the small dot in the vicinity of this
SU(4) fixed point. In Sections V and VI, we compare our predictions for transport
with other regimes where two channel Kondo (spin or charge) physics has been predicted. 
Section VII is devoted to the discussion of our results and mostly
we summarize our main experimental predictions for such a setup.
Technical details concerning the mapping onto the SU(4) fixed point are shown in appendix A
while the appendix B contains an analysis by conformal field theory of the vicinity of
the SU(4) fixed point.

\section{Model description} \label{sec:model}

In order to model the setup depicted in Fig. 1, 
we shall consider the Anderson-like Hamiltonian:
\begin{eqnarray}
\label{anderson}
H&=&\sum_{k\mu\sigma} \epsilon_k a^\dag_{\mu k\sigma} a_{\mu k\sigma}+
\sum_{p\sigma} \epsilon_p a^\dag_{p\sigma} a_{p\sigma} + 
{\hat Q^2\over 2C}+\varphi \hat Q \nonumber
\\ 
&+& \sum_{\sigma} \epsilon a_{\sigma}^{\dag}a_{\sigma}+U 
n_{\uparrow}n_{\downarrow}
\\ \nonumber
&+& t_{\mu}\sum_{k\mu\sigma}\left(a^\dag_{\mu k\sigma}a_{\sigma}+h.c.\right)+
    t_g\sum_{p\sigma}\hbox{\huge{(}}a^\dag_{p\sigma}a_{\sigma}+h.c.
\hbox{\huge{)}},
\end{eqnarray}
where $a_{\mu k\sigma}$, $a_{\sigma}$, $a_{p\sigma}$ are the annihilation
operators for electrons of spin $\sigma$ in the lead $\mu=1,2$, the small dot, and
the grain, respectively. Here, $t_{\mu}$ is the tunneling matrix element between the small dot and the lead $\mu$ which we assume  to be $k$ independent and $t_g$ denotes the
tunneling matrix element between the small dot and the grain. 

Moreover, $\hat Q$ denotes the charge operator of the grain such that $\langle \hat{Q}\rangle=eN_{grain}$, $C$ represents the capacitance between the grain and the plunger gate, and $\varphi$ is related to the back-gate voltage $V_g$ through $\varphi=-V_g$. 
$\epsilon<0$ and $U$ are respectively the energy level and charging energy of the small dot, and $n_{\sigma}=a_{\sigma}^{\dag}a_{\sigma}$ such that $N_{dot}=\sum_{\sigma} \langle n_{\sigma}\rangle$.

The direct inter-dot capacitive coupling is assumed 
to be weak and will be therefore neglected.

 We also assume that the grain is embodied by
a dense energy spectrum which implies that the grain is large enough such that its level spacing $\Delta_g$ is very small compared to its charging energy $E_c=e^2/(2C)$: $\Delta_g/E_c\rightarrow 0$.
We like mention that in the recent experiment of Ref. \onlinecite{David}, $E_c\approx 100\mu eV \approx 1K$ and $\Delta_g\approx 2\mu eV\approx 25mK$.

It is now convenient to introduce the linear combinations of the electron
operators 
\begin{equation}
\left(\begin{array}{c}
a_{k\sigma} \\
b_{k\sigma} 
\end{array}\right)
=\left( \begin{array}{cc}
\cos \alpha & \sin \alpha \\
-\sin\alpha & \cos \alpha
 \end{array} \right) 
 \left(\begin{array}{c}
a_{1k\sigma}  \\
a_{2k\sigma} 
\end{array} \right),
\end{equation}
\\
where the angle is determined by 
\begin{equation}
\tan \alpha=\frac{t_2}{t_1}.
\end{equation}
The Hamiltonian thus takes
the block-diagonal form:
\begin{eqnarray}
H&=&\sum_{k\mu\sigma} \epsilon_k a^\dag_{\mu k\sigma} a_{\mu k\sigma}+
\sum_{p\sigma} \epsilon_p a^\dag_{p\sigma} a_{p\sigma} + 
{\hat Q^2\over 2C}+\varphi \hat Q \nonumber
\\ 
&+& \sum_{\sigma} \epsilon a_{\sigma}^{\dag}a_{\sigma}+U 
n_{\uparrow}n_{\downarrow}
\\ \nonumber
&+& t\sum_{k\sigma}\left(a^\dag_{k\sigma}a_{\sigma}+h.c.\right)+
    t_g\sum_{p\sigma}\hbox{\huge{(}}a^\dag_{p\sigma}a_{\sigma}+h.c.
\hbox{\huge{)}},
\end{eqnarray}
with $t=\sqrt{t_1^2+t_2^2}$. We finally converge to an Anderson model with a
level hybridizing with the electrons of the grain as well as those of an {\it effective} reservoir lead. 

In Secs. \ref{sec:valent},  \ref{sec:transport} 
and \ref{sec:blockade}, we focus on the situation where the small dot 
resides in the Kondo regime which requires  the last level to be singly  occupied and the condition
\begin{equation}\label{Kondo}
(t,t_g)\ll (-\epsilon,U+\epsilon),
\end{equation}
to be fulfilled $(\epsilon<0)$. Moreover, for the small dot in the Kondo realm, one generally gets $U\gg E_c$; in Ref. \onlinecite{David}, $U\approx 1meV$. The mixed-valence limit of the small dot discussed in Refs. \onlinecite{Gramespacher,Schiller} will be commented in Sec. \ref{sec:2ckcharge}.

\section{SU(4) Kondo entanglement}\label{sec:valent}

In the local moment regime for the small dot, we can integrate out charge fluctuations in the small dot resorting to a generalized Schrieffer-Wolff transformation.

\subsection{Generalized Schrieffer-Wolff transformation}

In the vicinity of the degeneracy point  $\varphi=-e/2C$,
 where the grain charging states with $Q=0$ and $Q=e$ are degenerate implying that $N_{grain}=\langle Q\rangle/e = 1/2$ in Fig. \ref{phasediagram}, in the same spirit as a small dot coupled to two macroscopic electron leads, the system is embodied by 
\begin{eqnarray}
\label{ham0}
H&=&\sum_{k} \epsilon_k a^\dag_{k} a_{k}+
\sum_p \epsilon_{p} a^\dag_{p}a_{p} + 
{\hat Q^2\over 2C}+\varphi \hat Q\\ \nonumber
&+&\sum\limits_{m,n} \left(
{J_{mn}\over 2}\vec S\cdot{\vec{\sigma}}+V_{mn}\right) a_{m}^\dag 
a_{n}.
\end{eqnarray}
To simplify notations, the spin indices have been omitted and
hereafter. Here, $m,n$ take values in the two
sets ``effective lead'' (k) or ``grain'' (p), the spin $\vec S$ is the spin 
of the small dot, $\vec{\sigma}$ are Pauli matrices acting on the spin space
of the electrons in the reservoirs and in the grain. Let us now discuss the parameters $J_{mn}$ and $V_{mn}$ in more detail. The Schrieffer-Wolff transformation leads to
\begin{eqnarray}
J_{kk} &=& 2t^2\left[\frac{1}{-\epsilon}+\frac{1}{U+\epsilon}\right]\\ \nonumber
J_{pp} &=& 2t_g^2\left[\frac{1}{-\epsilon}+\frac{1}{U+\epsilon}\right]\\ \nonumber
J_{pk} &=& 2 t t_g \left[\frac{1}{-\epsilon}+\frac{1}{U+\epsilon}\right],
\end{eqnarray}
as well as
\begin{eqnarray}
V_{kk} &=& \frac{t^2}{2}\left[\frac{1}{-\epsilon}-\frac{1}{U+\epsilon}\right]\\ \nonumber
V_{pp} &=& \frac{t_g^2}{2}\left[\frac{1}{-\epsilon}-\frac{1}{U+\epsilon}\right]\\ \nonumber
V_{pk} &=& \frac{t t_g}{2} \left[\frac{1}{-\epsilon}-\frac{1}{U+\epsilon}\right].
\end{eqnarray}
It is relevant to note that $V_{mn}=0$ at the electrostatic symmetric point $U=-2\epsilon$ for the small dot.

We now make use of this set of couplings to determine when a SU(4) Kondo fixed point can be stabilized. This takes place for the symmetric tunneling condition,
\begin{equation}
t=\sqrt{t_1^2+t_2^2}=t_g,
\end{equation} 
which implies $J_{pp}=J_{kk}=J_{pk}\equiv J$ and $V_{pp}=V_{kk}=V_{pk}\equiv {\cal V}$. More precisely, the Hamiltonian becomes
\begin{eqnarray}
\label{ham}
H&=&\sum_{k} \epsilon_k a^\dag_{k} a_{k}+
\sum_p \epsilon_{p} a^\dag_{p}a_{p} + 
{\hat Q^2\over 2C}+\varphi \hat Q\\ \nonumber
&+&\sum\limits_{m,n} \left(
{J\over 2}\vec S\cdot{\vec{\sigma}}+{\cal V}\right) a_{m}^\dag 
a_{n}.
\end{eqnarray}
This Hamiltonian has been previously shown to exhibit a SU(4) Kondo Fermi liquid type fixed
point.\cite{Karyn} 

\subsection{SU(4) Hamiltonian for $t\approx t_g$}

The main point is as follows. Close to the degeneracy point $\varphi=- e/2C$ and for $k_B T\ll E_c$, only
the states with $Q=0$ and $Q=e$ are accessible and higher energy states 
can be removed from our theory introducing 
the projectors $\hat P_0$ and $\hat P_1$ (which project on the states 
with $Q=0$ and $Q=e$ in the grain, respectively). The truncated
Hamiltonian (\ref{ham}) then takes the following form:
\begin{eqnarray}
H&=&\sum\limits_{k,\tau=0,1} \epsilon_k 
a_{k\tau}^\dag a_{k\tau}\left(\hat P_0+\hat P_1\right)+e\tilde{h}\hat P_1\\ 
\nonumber
&+& \sum\limits_{k,k'}\hbox{\Huge{[}}\left({J\over 2}{\vec \sigma}\cdot\vec S
+V\right)\left(a_{k1}^\dag a_{k'0}\hat P_0+a_{k'0}^\dag a_{k1}\hat P_1\right)
\\ \nonumber
&+&\sum_{\tau=0,1} \left({J\over 2}{\vec\sigma}\cdot\vec
S+{\cal V}\right)a_{k\tau}^\dag a_{k'\tau}\hbox{\Huge{]}},
\end{eqnarray}
where now the index 
$\tau=0$ indicates the reservoir leads (referring to the $a_{k\sigma}$ operators) and $\tau=1$ indicates the grain. We have also introduced the small parameter 
\begin{equation}
\tilde{h}=\frac{e}{2C}+\varphi=\frac{e}{2C}-V_g\ll \frac{e}{C},
\end{equation}
which measures deviations from the degeneracy point. Remember that $\tau$ can be viewed as an abstract orbital degree of freedom and the Hamiltonian can be rewritten in an SU(4) form by introducing another set of Pauli matrices for the orbital sector. For sake of clarity, technical details have been shown separately in Appendix A.

At low energy, for $\tilde{h}=0$, the system is described by an SU(4) {\it symmetrical} Kondo interaction of the form:\cite{Karyn}
\begin{eqnarray} \label{irrrep}
\hskip -1.2cm
H_K \hskip -0.15cm &=& \hskip -0.15cm 
J \sum\limits_{A;i,j} \psi^{\dag}_{i} T^A_{ij}
\left[\sum\limits_{\alpha\beta}\left(S^{\alpha}+{1\over 2}\right)
\left(T^{\beta} +{1\over 2}\right)\right]^A\psi_{j} \\ \nonumber
 \hskip -0.15cm &=&  \hskip -0.15cm {J\over 4}
\sum\limits_{A} M^A\sum\limits_{i,j} \psi^{\dag}_{i}
T^A_{ij}\psi_{j}.
\end{eqnarray}
The Kondo effect is local: $\psi_i=\sum_{k\tau} a_{k\tau}$ and $a_{k\tau}$ is defined in Eq. (11).
We have introduced the ``hyper-spin'' 
\begin{equation}
M^A\in \left\{2S^{\alpha},2T^{\alpha},
4S^{\alpha}T^{\beta}\right\},
\end{equation}
for $\alpha,\beta=x,y,z$. 
The electron operator $\psi$ transforms under the fundamental 
representation of the SU(4) group, with generators $T^A_{ij}$  
$(A=1,...,15)$, and the index $i$ labels the four combinations of
possible spin $(\uparrow,\downarrow)$ 
and orbital indices $(0,1)$, which means $(0,\uparrow)$, $(0,\downarrow)$,
$(1,\uparrow)$ and $(1,\downarrow)$.  Note that the lead operators $a_{k\sigma}$ appear
explicitly as electron operators of the SU(4) theory through $\psi_{0,\sigma}=\sum_k a_{k\sigma}$.
Remember also that here $\vec{S}$ denotes the spin of the small dot in the
Kondo regime whereas $T^z$ measures the charge on the large grain close to a degeneracy
point, namely $Q = e \left({1\over 2} + \langle T^z \rangle\right)$, and $T^+|Q=0\rangle = |Q=1\rangle$
as well as $T^-|Q=1\rangle = |Q=0\rangle$ mimic the charge flip associated to the transfer of electrons from lead to grain or vice-versa. The emergence of the
strongly-correlated SU(4) ground 
state, described by the screened hyper-spin
\begin{equation}
\left(\vec{S}+{1\over 2}\right)
 \left(\vec{\hbox{T}}+{1\over 2}\right),
\end{equation}
must be clearly attributed to the 
strong {\it entanglement} 
between the {\it charge} degrees of freedom of
the {\it grain} and the {\it spin} degrees of freedom of the {\it small dot} 
at low energy induced by the prominence of spin-flip assisted tunneling between the leads and
the grain via the small dot.

 It is important to keep in mind that at the SU(4) Kondo fixed point, 
both the spin $\vec{S}$ and the orbital spin $\vec{T}$ are screened; note that
$\langle T^z\rangle =0$ for $h=0$ that is in accordance with the fact that $N_{grain}=1/2$ at $\varphi=-e/2C$.

At this stage, we like to emphasize
that due to the large emergent symmetry in the problem, the Kondo scale
\begin{equation}
k_B T_K^{SU(4)}\sim D\ e^{-1/4\nu J},
\end{equation}
is much larger than those of Kondo problems with an underlying SU(2) symmetry; $\nu$ is the constant density of states obtained after linearization of the electron dispersion relation and $D\sim \min(E_c,\Delta_d)$ with $\Delta_d$ being the level
spacing in the small dot (from Ref. \onlinecite{David}, one estimates $\Delta_d\approx E_c\approx 100\mu eV\approx 1K$). Experimental indications of this feature have been recently reported in Ref. \onlinecite{KG}.

Furthermore, in contrast to the  two-channel Kondo fixed point which is known to be extremely fragile
with respect to relatively small perturbations (like a channel asymmetry or the application of a magnetic field), the SU(4) Kondo fixed point which yields a Fermi liquid behavior is robust at least for weak perturbations. This point will be clearly emphasized in the next section.

\section{Linear transport}\label{sec:transport}

Let us now discuss explicitly the transport between the lead 1 (drain) and lead 2 (source).  Along the lines of Ref. \onlinecite{Pustilnik}, for a small bias voltage $V\rightarrow 0$ between source and drain, it is straightforward to write down the related conductance through the Kubo formula,
\begin{equation}
G= \frac{G_0}{2}\sum_{\sigma=\uparrow,\downarrow} \int d\omega\ \left(-\frac{d f}{d\omega}\right)
\times (-\hbox{Im}\left[\pi\nu{\cal T}_{\sigma}(\omega,T)\right]),
\end{equation}
where 
\begin{equation}
G_0=\frac{2e^2}{h}\sin^2 (2\alpha).
\end{equation}
Throughout the text, we will set $\hbar=1$.
In the formula above, $f(\omega)$ is the Fermi function ($\omega$ is the energy measured from the Fermi level) and ${\cal T}_{\sigma}$ is the T-matrix for the $a_{k\sigma}$ particles.\cite{Leonid} The ground states of the various Kondo models discussed in this section are embodied by the formation of singlets from the spin sector and also possibly from the orbital sector, and hence are not degenerate. This results in the important fact that the T-matrix of the $a_{k\sigma}$ electrons is completely characterized by the scattering phase shifts $\delta_{\sigma}$ at the Fermi level through\cite{Newton,Leonid}
\begin{equation}\label{tmatrix}
-\pi {\cal T}_{\sigma}(\omega=0,T=0)=\frac{1}{2i}\left(e^{2i\delta_{\sigma}}-1\right),
\end{equation}
or
\begin{equation}
G(V=0,T=0) =G_0 \sum_{\sigma=\uparrow,\downarrow} \frac{\sin^2\delta_{\sigma}}{2}.
\end{equation}

\subsection{$\frac{\pi}{4}$ phase shift}

 Interestingly, as seen above, the energy spectrum of the SU(4) fixed point can be understood as a sum of four independent spinless fermions with phase shifts $\delta_i=\pi/4$. This phase shift is characteristic of the SU(4) Hamiltonian and simply emerges from the Friedel sum rule $1=\sum_{i=1}^4 \delta_i/\pi$ ($\sum_{i=1}^4 \delta N_i=1$ being the total number of electrons to compensate for the magnetic moment of the small dot);\cite{Borda} the main point being that two of these four electrons embody the electrons of the leads 1 and 2 which ensures $\delta_{\sigma}=\pi/4$ in Eq. (\ref{tmatrix}). We hence predict:
 \begin{eqnarray}
 G(V=0,T=0) &=& \frac{G_0}{2}=\frac{e^2}{h}\sin^2 (2\alpha)\\ \nonumber
 &=&\frac{e^2}{h}\frac{4(t_1 t_2)^2}{\left(t_1^2+t_2^2\right)^2}.
 \end{eqnarray}
 
It is worth noting the main difference with the
 SU(2) case where one rather estimates $\delta N_{\uparrow} - \delta N_{\downarrow} =1$. Moreover, exploiting the Friedel sum rule  $\delta N_{\sigma}=\delta_{\sigma}/\pi$ and particle-hole symmetry 
 $a_{k\sigma}\rightarrow \sigma a^{\dagger}_{k-\sigma}$ (or $\delta_{\uparrow}+\delta_{\downarrow}=0$) this rather results in $\delta_{\sigma}=\sigma \pi/2$ and $\sigma=\pm$ for $\sigma=\uparrow,\downarrow$, respectively.\cite{Leonid} In the case of the SU(4) Kondo fixed point, the fact that $\delta_{\uparrow}+\delta_{\downarrow}\neq 0$ translates the breaking of the particle-hole symmetry as can be seen from Eq. (10); {\it e.g.}, the potential scattering terms $V_{pk}$ are strongly relevant. This will affect transport properties accordingly.

 One signature of the SU(4) symmetry in the linear transport thus corresponds to
 a conductance which is always smaller or equal than $e^2/h$ inspite of the complete screening of the
 local moment of the small dot. This feature makes the predicted SU(4) symmetry in our geometry  already distinguishable
from the quantum dot carbon nanotube geometry where some SU(4) symmetry may have been observed.\cite{KG}
 It is indeed extremely difficult to differentiate between an ordinary two-level SU(2) Kondo effect and the SU(4) Kondo state when based only on linear transport as a result of the two possible chiralities (clockwise or counter-clockwise) in the nanotube.\cite{Rosa1,Rosa2} We also judge that our geometry in a certain sense is more convenient than that suggested in Ref. \onlinecite{Borda} where the Kondo SU(4) fixed point requires an infinitely small coupling between the small dots and hence will result in a tiny zero-bias conductance peak when considering transport accross the double dot structure.
 
 On the other hand, a halved conductance $\sim e^2/h$ is also characteristic of two-channel Kondo 
 fixed points and thus to distinguish the SU(4) Kondo fixed from other Kondo candidates
 we shall examine corrections at finite temperature or finite bias. Moreover, for a setup with quite asymmetric values of $t_1$ and $t_2$, the zero-bias conductance will be always much weaker than $2e^2/h$, even for an ordinary Kondo effect, and thus the value of the conductance at $V=0$ and $T=0$ might not be very meaningful.\cite{David}
 
 \subsection{Exotic Fermi-liquid corrections}
 
The SU(4) fixed point exhibits a Fermi-liquid like behavior and the corrections to the zero-bias conductance can be {\it a priori} computed from Nozi\`eres\cite{No} Fermi liquid theory.
We have explicitly computed these corrections using a conformal field theory approach following
Affleck and Ludwig. \cite{AL} The details of this analysis  are reported in the appendix \ref{sec:cft}.
At finite temperature, we estimated
 \begin{equation}\label{FLcorrection}
 G(V=0,T)-\frac{G_0}{2} \propto -G_0\left(T/T_K^{SU(4)}\right)^3,
 \end{equation}
where we have assumed that $T\ll T_K^{SU(4)}$. In contrast to the standard $SU(2)$ case,
governed by the scaling law,\cite{Leonid}
\begin{equation}
G(V=0,T)=G_0\left[1-\left(\pi T/T_K^{SU(2)}\right)^2\right],
\end{equation}
where $k_B T_K^{SU(2)}\sim D\ e^{-1/\nu J}$, we emphasize that
the temperature corrections to the conductance are much weaker here.
This scaling law is another feature of the SU(4) fixed point which is clearly different from
the SU(2) case or some two channel Kondo fixed point. The absence of a quadratic term in temperature
inspite of the Fermi liquid nature of the Kondo fixed point much be attributed to the extra (purely imaginary) factor $e^{i2\delta}=i$ --- that reflects the incipient SU(4) symmetry in the infra-red ---
entering in the expression of the inelastic contribution.\cite{Leonid}

Moreover, at temperature $T=0$, the conductance evolves linearly with the bias
voltage. More precisely, for low bias voltage $eV\ll k_B T_K^{SU(4)}$ we found that
 \begin{equation}
G(V,T=0)\approx\frac{G_0}{2}\left(1+\frac{15 \pi}{2}\frac{eV}{k_BT_K^{SU(4)}}\right).
 \end{equation}
Such a linear behavior of $G(V,T=0)$ stems from the absence of particle-hole symmetry 
 $a_{k\sigma}\rightarrow \sigma a^{\dagger}_{k-\sigma}$  due to, {\it e.g.}, the relevant potential scattering terms $V_{pk}$ and $V_{kp}$. More precisely, in the SU(4) Kondo model the Abrikosov-Suhl resonance is located at $\omega\sim k_B T_K^{SU(4)}$ above the Fermi energy.\cite{Rosa1} The SU(4) Kondo fixed point can be thus viewed as a fictitious level with a density of states:
\begin{equation}\label{rho_omega}
\rho(\omega)\approx\frac{1}{\pi}\frac{k_B T_K^{SU(4)}}{\left(\omega-k_B T_K^{SU(4)}\right)^2+\left(k_B T_K^{SU(4)}\right)^2}.
\end{equation}
Note that the SU(4) Kondo temperature appearing in Eq. (\ref{rho_omega}) may slightly differ from the SU(4) Kondo temperature extracted from the Fermi liquid analysis.
The conductance accross the small dot at low temperatures
should approximately follow $\rho(\omega=eV)$. In the linear regime $eV\ll k_B T_K^{SU(4)}$ one
recovers the linear dependence with the bias voltage which was predicted using the more rigorous
conformal field theory approach:
\bea
G(V,T=0)&\sim& G_0 \rho(\omega)\pi k_B T_K^{SU(4)} \\
&\sim& \frac{G_0}{2} \left(1+{\cal O}\left(\frac{eV}{k_B T_K^{SU(4)}}\right)\right).\nn
\eea
Here, $\rho(\omega)\pi k_B T_K^{SU(4)}$ mimics the effective transmission probability accross the small dot at the energy $\omega$ and $\pi k_B T_K^{SU(4)}$ being the effective energy bandwidth in the Kondo problem. This is clearly distinguishable from the standard one-channel SU(2) Kondo effect. We argue that the absence of a zero-bias peak in the conductance is the most significant hallmark of the SU(4) Kondo realm. Moreover, for the SU(4) Kondo state, a peak in the differential conductance $dI/dV$ might rather appear around $eV\sim k_B T_K^{SU(4)}$ where $\rho(eV)$ is maximum.
 
The fixed point Hamiltonian also allows one to calculate the corrections to the conductance due to a finite magnetic field B. One can check that inside the SU(4) Kondo realm, the conductance has no quadratic contribution in the magnetic field inspite of the Fermi liquid type fixed point. More precisely, along the lines of Ref. \onlinecite{No}, for small $g\mu_B B\ll k_B T_K^{SU(4)}$, the phase shifts of the $a_{k\sigma}$ take the form (we use $\sigma=\pm$ for $\sigma=\uparrow,\downarrow$)
\begin{equation}
\delta_{\sigma}(B)=\frac{\pi}{4}+\sigma \frac{g\mu_B B}{k_B T_K^{SU(4)}}.
\end{equation}
This immediately result in
 \begin{equation}
 G(V=0,T=0,B)\approx\frac{G_0}{2}.
  \end{equation}
Thus, for $g\mu_B B\ll k_B T_K^{SU(4)}$,  there is no second order correction in $B$ to the conductance at zero temperature. At this point, one might legitimately wonder how the conductance evolves for stronger magnetic fields. To answer this question, let us apply a strong magnetic such that spin Kondo physics with the small dot is suppressed.\cite{Karyn} In this case, orbital degrees of freedom $V_{\perp}$ and $V_z$ can still develop a purely orbital one-channel Kondo effect through the screening of the orbital pseudo-spin of the grain $\vec{\hbox{T}}$ (see Appendix \ref{sec:mapping}). This will result in $\delta_{\uparrow}=\pi/2$ whereas $\delta_{\downarrow}=0$.\cite{Borda}  We infer that the $T=0$ conductance should still remain equal to $G_0/2$. However, we should mention that the emergent Kondo temperature $T_K[B=\infty]$ is much smaller than $T_K^{SU(4)}$; see Eq. (A11).

 \subsection{Channel asymmetry}
 
 \begin{figure}[ht]
\centerline{\epsfig{file=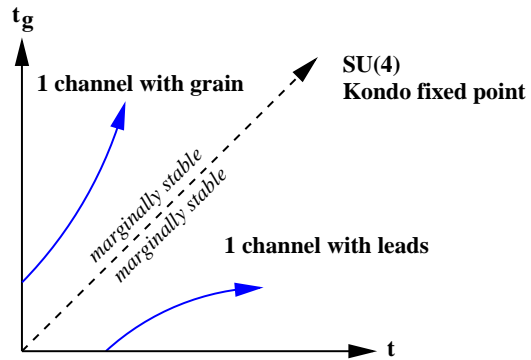,angle=0.0,height=4.7cm,width=7cm}}
\caption{(color online) Fixed points for a grain at degeneracy. The SU(4) fixed point requires 
$t\approx t_g$ and is marginally stable. For a large asymmetry between $t=\sqrt{t_1^2+t_2^2}$ and
$t_g$ the small dot will undergo a crossover towards a single-channel Kondo effect either with the reservoir leads or with the grain.}
\label{flow}
\end{figure}
 
 Now, we shall discuss the effect of a small channel asymmetry. The Numerical Renormalization Group (NRG) approach clearly emphasizes that a small channel anisotropy $\delta J=(J-J_{pp})$ does not hinder the SU(4) Kondo fixed point to develop.\cite{Karyn} Moreover, as summarized in Appendix B, the effect of a small channel asymmetry is (almost) equivalent to the application of an orbital magnetic field. 
 Note that $\delta J$ can be controlled, {\it e.g.}, through the tunneling amplitude $t_g$. 
Now, let us  modify the tunneling amplitude $t_g$ to produce a finite channel asymmetry; $G_0$ remaining fixed and thus $J_{kk}=J$. The SU(4) Kondo realm will remain marginally stable as long as $|\delta J|<k_B T_K^{SU(4)}$ which implies that the zero-bias conductance peak at $G_0/2$ will be still detectable for small deviations from $\delta J=0$. We predict the smooth evolution
 \begin{equation}
 \delta_{\sigma}\sim\frac{\pi}{4} + \frac{\delta J}{T_K^{SU(4)}},
 \end{equation} 
 and hence
 \begin{eqnarray}
 G(V=0,T=0) &=& G_0\sin^2\left(\frac{\pi}{4} + \frac{\delta J}{T_K^{SU(4)}}\right) \\ \nonumber
 & \approx & \frac{G_0}{2}\left(1 + \frac{2\delta J}{T_K^{SU(4)}}\right).
 \end{eqnarray}
 
 For a strong channel anisotropy  such that for $|\delta J|\gg k_B T_K^{SU(4)}$ the small dot will undergo a single-channel Kondo effect. This is summarized in Fig. \ref{flow}. The orbital degrees of freedom are somehow quenched by the strong channel asymmetry\cite{note1} and only the spin Kondo physics will rule out the linear transport accross the small dot.
  
 For $\delta J\ll 0$, the spin Kondo coupling $J_{pp}$ between
 the grain and the small dot will be the largest one through the Renormalization Group flow\cite{Karyn} leading to an SU(2) Kondo effect with $\delta_{\sigma}=0$ whereas for $\delta J\gg 0$, the spin Kondo coupling $J_{kk}=J$ between the small dot and the leads will rule out
 the low-energy physics leading to a screening of the small dot's spin by the lead electrons, {\it i.e.}, to $\delta_{\sigma}=\sigma\pi/2$, and thus for $k_B T\ll k_B T_K^{SU(2)} \sim De^{-1/\nu J}$ the conductance will exhibit a plateau at the maximum value $G_0$ allowed by quantum mechanics.
 
 Remember that the conductance evolves smoothly from $G_0/2$ either to zero or to $G_0$ dependently on the sign of $\delta J$. This seems {\it a priori} conceivable to observe experimentally those interesting predictions by continuously tuning the tunnel barrier amplitude $t_g$.

\section{Coulomb blockaded grain}\label{sec:blockade}

Now, let us start with the SU(4) fixed point which again requires the condition $t\approx t_g$ and progressively move away from the degeneracy point of the grain $\varphi=-e/2C=-V_g$ such that $\tilde{h}=\varphi+e/2C> 0$. In our orbital representation, $\tilde{h}$ can be identified as a local orbital magnetic field acting on the orbital spin $\vec{\hbox{T}}$; see Eq. (A1).

\subsection{Charging energy in the grain}

 A weak local orbital field $e\tilde{h}\ll k_B T_K^{SU(4)}$ will not destroy the marginally stable SU(4) fixed point. On the other hand, similar to the channel asymmetry discussed precedingly strong deviations from the degeneracy point of the grain $\varphi=-e/2C$ will completely hamper the orbital Kondo screening to take place. This can be explicitly understood as follows. The grain becomes Coulomb blockaded and as a result it costs a finite energy $E_{c-1}=E_c(1+2N)>0$, where $N=C V_g/e < 1/2$ and $N_{grain}=0$, to add a {\it hole} on the grain. In a similar way, it costs $E_{c1}=E_c(1-2N)>0$ to add an {\it electron} onto the grain. The lead-dot and grain-dot Kondo couplings, $J_{kk}$ and $J_{pp}$ respectively, 
then become {\it asymmetric} for $t = t_g$:
\begin{eqnarray}\label{couplingss}
J_{kk} &=& 2t^2\left[\frac{1}{-\epsilon}+\frac{1}{U+\epsilon}\right]=J\\ \nonumber
\bar{J}_{pp} &=& 2t_g^2\left[\frac{1}{E_{c1}-\epsilon}+\frac{1}{U+\epsilon+E_{c-1}}\right].
\end{eqnarray}
The virtual intermediate state where an electron first
hops from the grain onto the small dot induces
an excess of energy $E_{c-1}$. The first term in the second equation contains
the energy of the intermediate state of the process where the temporal order 
of the hopping events is reversed. Moreover, in the temperature range $k_B T< (E_{c1},E_{c-1})$, the off-diagonal processes where, {\it e.g.}, an electron from the a reservoir lead flips the impurity spin of the small dot and then jumps onto the grain are suppressed exponentially as 
$\bar{J}_{kp}(T)\approx \bar{J}_{kp}e^{-E_{c1}/4k_BT}$, where in the Coulomb blockaded grain regime we obtain
\begin{equation}
\bar{J}_{kp}=2 t t_g\left[\frac{1}{E_{c1}-\epsilon}+\frac{1}{U+\epsilon}\right],
\end{equation}
whereas the diagonal spin processes $J_{kk}$ and $\bar{J}_{pp}$ can be strongly renormalized at low temperatures. In other words, in
the Renormalization Group language, if we start at high temperature with a
set of Kondo couplings $J_{kk}, \bar{J}_{pp}, \bar{J}_{kp}, \bar{J}_{pk}$, the growing of 
$\bar{J}_{kp}, \bar{J}_{pk}$ is cut-off when $k_B T$ is decreased below $\hbox{max}(E_{c1},E_{c-1})$ 
whereas the growing of $J_{kk}, \bar{J}_{pp}$ is not. The other off-diagonal term $\bar{J}_{pk}$ where an electron hops from the grain flips the dot's spin and hence jumps onto one reservoir lead reads
\begin{equation}
\bar{J}_{pk}=2 t t_g\left[\frac{1}{-\epsilon}+\frac{1}{U+\epsilon+E_{c-1}}\right].
\end{equation}
Following the same reasoning, the potential scattering terms $\bar{V}_{pk}$ and $\bar{V}_{kp}$ become also suppressed exponentially.

Thus, starting with the SU(4) condition $t_g=t$ and progressively enhancing the orbital magnetic field by tuning $V_g$ will generally produce the quenching
of the dot's spin by the lead electrons leading to a single-channel (spin) Kondo effect and consequently to $G(T=0,V=0)=G_0$.  

\subsection{2-channel spin Kondo physics}

Now, it is important to underline that this also offers an opportunity to reach a 2-channel Kondo effect in the spin sector (denoted 2CK spin in Fig. \ref{phasediagram}) for asymmetric tunneling junctions provided the condition $J_{kk}=\bar{J}_{pp}$ can be reached with a fine-tuning of the 
gate voltages.\cite{oreg} For instance, at the particle-hole condition both for the dot and the grain requiring $U=-2\epsilon$ and $E_{c1}=E_{c-1}=E_c$ the 2-channel Kondo effect should roughly emerge when $t_g/t\approx 1+E_c/U$ assuming $E_c/U\ll 1$. In Ref. \onlinecite{David}, one extracts $E_c/U\approx 10^{-3}\rightarrow 0$, and thus
 the 2-channel spin Kondo effect and the SU(4) Kondo correlated liquid might be still connected by
 varying $N_{grain}$ as shown in Fig. \ref{phasediagram}. The nature of the transition between
 those two Kondo candidates is beyond the scope of this paper.
 
 We like to emphasize that the observation of a 2-channel spin Kondo effect represents a challenging goal on its own due to its extreme fragility. Let us remind that the two-channel Kondo realm is embodied by a zero-bias peak of height $G_0/2$;\cite{oreg,Pustilnik} on the other hand, the detection of such a plateau in the Coulomb blockaded grain realm represents a genuine challenging experimental task on its own in the sense that a small deviation from the ideal point $J_{kk}=\bar{J}_{pp}$ will produce
either a unitary conductance $G_0$ or zero at low temperatures dependently on the sign of $(J_{kk}-\bar{J}_{pp})$. This stands for the fundamental difference between the 2-channel (spin) fixed point and the SU(4) one; the latter is marginally robust and is subject to smooth crossovers whereas the former is unstable towards small perturbations.\cite{Pustilnik} Another difference is the temperature-correction to the zero-bias peak and more precisely for the 2-channel Kondo realm,  one expects\cite{Pustilnik}
\begin{equation}
G(V=0,T)=\frac{G_0}{2}\left(1-\sqrt{\pi T/T_K^{2CKs}}\right),
\end{equation}
which is very different from the SU(4) or SU(2) scaling corrections. The 2-channel (spin) Kondo temperature $T^{2CK}_K$ is of the order of $T_K^{SU(2)}$. The $V$-scaling
reads
\begin{equation}
G(V,T=0)=\frac{G_0}{2}\left(1-\frac{3}{\sqrt{\pi}}\sqrt{|eV|/k_BT_K^{2CKs}}\right).
\end{equation}
Such a scaling has been recently evidenced in Ref. \onlinecite{David}

\section{2-channel orbital Kondo physics}\label{sec:2ckcharge}

In the previous sections, the small dot was assumed in the Kondo regime. One may wonder what happens when the dot gate voltage is tuned towards the mixed valence regime where $N_{dot} \sim 1/2$. The spin is no longer a quantum number and the Kondo effect is destroyed. Let us analyze
the resonant level situation where $\epsilon\sim 0$. The grain will be at the degeneracy condition $\varphi=-e/2C$. We briefly rehearse the main lines of Ref. \onlinecite{Schiller}. 
When $\epsilon\sim 0$, resonant tunneling occurs through the small dot and we are mainly interested
in the influence of the grain on the resonant transport. We introduce
the two relevant energy scales $\Gamma_t=\pi \sum_k t^2 \delta(\epsilon_k)$ and $\Gamma_g=\pi\sum_p t_g^2 \delta(\epsilon_p)$ corresponding to half the tunneling rates from the
small dot to the effective lead and to the grain, respectively.\cite{Schiller}

The key point is that the conduction leads and the small dot can be formally replaced by a single lead  with a huge renormalized density of states at the Fermi level,
\begin{equation}
\rho_{eff}(\epsilon)=\frac{1}{\pi\Gamma_t}\theta\left(\frac{\pi\Gamma_t}{2}-|\epsilon|\right)=
1/(\pi\Gamma_t).
\end{equation}
Close to the degeneracy point $\varphi=-e/2C=-V_g$ of the grain, one can still introduce the orbital
spin $\vec{T}$ that embodies the two allowed charge states $Q=0$ and $Q=e$ in the grain. Transferring electrons from the effective lead with the density of states $\rho_{eff}$ onto the grain and vice-versa then results in the 2-channel ``charge'' Kondo problem originally introduced by Matveev;\cite{Matv1} see Appendix A. The question whether 2-channel Kondo physics through extra orbital degrees of freedom could eventually arise for a small dot at resonance then naturally emerges. 

For $\Gamma_g\ll \Gamma_t$ $(t_g\ll t)$, the authors of Ref. \onlinecite{Schiller} have established that the two-channel orbital Kondo scale is extremely small $k_B T_K^{2CKo}=\sqrt{E_c \Gamma_g} \exp[-\pi^2/2\sqrt{{\cal T}_g}]\ll E_c$ and therefore should not be detectable; ${\cal T}_g$ being proportional to an effective transmission coefficient through the grain ${\cal T}_g=4\Gamma_g/\Gamma_t$. Therefore, for $\Gamma_g\ll \Gamma_t$, the conductance at low temperatures should only slightly deviate from its maximum value $G_0$. On the other hand, when $\Gamma_g\rightarrow \Gamma_t$ meaning ${\cal T}_g\approx 4$, one should observe a noticeable increase of $k_B T_K^{2CKo}\sim \sqrt{E_c\Gamma_g}$ which might then become observable. If this is the case, the electron lead operators $a_{k\sigma}$ then will be subject to a $\pi/4$ phase shift producing  a zero-bias peak of height $G_0/2$ at low temperatures. This sudden variation of the zero-bias peak height should be unambiguously detectable in experiments by increasing $t_g$. The Kondo temperature 
$T_K^{2CKo}$ grows linearly with $\sqrt{\Gamma_g}$ for $\Gamma_g\gg\Gamma_t$.
Moreover, by applying a substantial magnetic field one predicts $\delta_{\uparrow}(B)\rightarrow \pi/2$ while $\delta_{\downarrow}(B)\rightarrow 0$ and thus the conductance height should remain 
of the order of $G_0/2$ similar to the SU(4) Kondo effect. 

This 2-channel orbital Kondo effect should be then observable when\cite{Schiller} $|\epsilon|\ll \Gamma_t$ (at the resonant peaks of the small dot) assuming that $T_K^{2CKo}\sim \sqrt{E_c\Gamma_g}$ is detectable while the SU(4) Kondo effect develops in the Coulomb valleys of the small dot with an odd number of electrons. 


 \section{Conclusion}

To summarize succintly, a bunch of exotic Kondo liquids other than the well-known SU(2) Kondo Fermi
liquid can possibly emerge in the mesoscopic structure with a small electron droplet (pistil or small dot) tunnel coupled to a larger metallic grain (petal) as well as to two macroscopic reservoir leads serving as source and drain. Let us mention that such a device has been realized recently.\cite{David}

Close to the degeneracy points of the grain and in a Kondo valley of the small dot (with a odd number of electrons) we have two spin objects hidden in the mesoscopic structure: the spin $\vec{S}$ of the last occupied level of the small dot as well as the orbital pseudo-spin $\vec{\hbox{T}}$ of the grain which mimics the two allowed charging states of the grain close to a given degeneracy point. As explained in the sequel, tunneling events between the grain and the reservoir leads involving (flipping) the spin of the
small dot becomes very prominent at low energy implying an obvious spin-orbital mixing at the
infra-red fixed point. The SU(4) group is the minimal group allowing such an orbital-spin entanglement
and which guarantees rotational invariance both in spin and orbital spaces. Some unambiguous consequences for the emergence of such an SU(4) Kondo state in this geometry are, {\it e.g.}, a ``halved'' zero-bias conductance accross the small dot at $T=0$. This behavior is very robust to any small temperature corrections which are in ${\cal O}(T^3)$ at $T\ll T_K^{SU(4)}$ or to the application
of a magnetic field. The absence of particle-hole symmetry, {\it e.g.}, due to the relevant
Matveev type scattering potential terms, reflects itself in an Abrikosov-Suhl resonance 
located at $\omega\sim k_B T_K^{SU(4)}$ and thus in a conductance that varies
linearly with $V$. 

The absence a zero-bias peak is the most significant hallmark of the
SU(4) liquid; one might expect a peak in $dI/dV$ at $eV\approx k_B T_K^{SU(4)}$.

We also want to stress that the SU(4) Kondo realm is marginally robust towards a small channel asymmetry. This SU(4) Kondo
domain should be comfortably detectable experimentally taking into account that the SU(4) Kondo energy scale should exceed that of the SU(2) Fermi liquid regime. Furthermore,  a substantial channel asymmetry or by pushing the large grain in the Coulomb blockade regime generally results in smooth crossovers towards a more conventional SU(2) Fermi liquid. The conductance should progessively saturate to the maximum value $G_0$ allowed by quantum mechanics or to $0$ dependently on the sign of the channel asymmetry. 

This geometry is definitely promising and also offers an opportunity to reach two different 2-channel
Kondo effects. For instance, let us start from the SU(4) Kondo state and progressively push the level
of the small dot at the resonant condition $\epsilon=0$ (instead of the Kondo valley for the SU(4) realm). Following Ref. \onlinecite{Schiller}, one could hence expect a two-channel orbital ``Matveev'' Kondo effect\cite{Matv1} with impressively a Kondo scale which should be accessible due to the huge density of states at the Fermi energy stemming from the resonant condition on the small dot. The linear conductance should remain tied at $G_0/2$; we have discussed how to distinguish experimentally between the SU(4) and the 2-channel Matveev fixed points. For instance, in contrast to the SU(4) Kondo state, the 2-channel orbital Kondo state should still arise when substantially increasing the grain-dot tunneling amplitude $t_g$ such that $t_g\gg t$; additionally, following Ref. \onlinecite{Schiller} one should observe an enhancement of the 2-channel Kondo energy scale. The stability of the 2-channel orbital state towards a substantial channel asymmetry $t_g\gg t$ stems from the important fact that in the Matveev Kondo problem, the two charge configurations in the grain play the role of the impurity spin whereas the physical spin of the conduction electrons acts as a passive channel index. A channel asymmetry in the Kondo couplings can only be driven through the application of an in-plane magnetic field.\cite{Georg} The precise nature of the quantum phase transition between the 2-channel orbital state and the SU(4) Kondo phase for $t_g\approx t$ when varying $\epsilon$ is an open question and goes beyond the scope of this paper. 

Finally, when the large grain becomes Coulomb blockaded (suppressing single-particle hopping events between the small dot and the grain) and the small dot lies in the Kondo valley, for meticulous fine-tunings of the different gate voltages a two-channel spin Kondo effect arises requiring that the grain-dot Kondo coupling is exactly equal to the (effective) lead-dot Kondo coupling.\cite{oreg,Pustilnik} Some experimental endeavors have been performed in this direction.\cite{David} However, this 2-channel spin Kondo fixed point is fragile towards any small perturbation. In Ref. \onlinecite{David}, one estimates
$E_c/U\sim 10^{-3}\rightarrow 0$ and thus the 2-channel spin Kondo effect and the SU(4) correlated liquid might be connected by varying $N_{grain}$ through $V_g$. 


{\it Acknowledgments: } The authors acknowledge stimulating discussions with D. Goldhaber-Gordon. K.L.H. thanks the hospitality of the Aspen center for Physics through the workshop on {\it Interactions, Coherence $\&$ Control in Mesosocopic Systems}. K.L.H. was supported by CIAR, FQRNT, and NSERC. P.S. has been partially supported by the contract PNANO-QSPIN of the ANR. D.L. acknowledges support from the Swiss NSF, the NCCR Nanoscience, ONR, and JST ICORP.

\appendix

\section{Mapping onto the SU(4) fixed point}\label{sec:mapping}

We find it appropriate to rehearse the main steps leading to the SU(4) fixed point again in the situation
where the small dot lies in the Kondo realm and the large grain is close to one degeneracy point, {\it
e.g.}, $\varphi=-e/2C$. 

In particular, for a large repulsion $U\gg E_c$ on the small dot, we note some
disagreement with Ref. \onlinecite{Schiller} which then will be discussed in detail in this Appendix.

\subsection{The bare Hamiltonian}

The full Hamiltonian (11) can be turned into\cite{Matv1,Georg}
\begin{eqnarray}
H&=&\sum\limits_{k,\tau} \epsilon_k a_{k\tau}^\dag a_{k\tau}+e\tilde{h}T^z 
\\ \nonumber
&+& \sum\limits_{k,k'}\hbox{\Huge{[}}     
\sum\limits_{\tau,\tau'}\left({J\over 2}{\vec\sigma}\cdot\vec S+{\cal V}\right)
\hbox{\Large{(}}\tau^x T^x+\tau^yT^y\hbox{\Large{)}}_{\tau,\tau'}
a_{k\tau}^\dag a_{k'\tau'}
\\ \nonumber
&+ &\hskip 0.7cm 
\sum_\tau\left({J\over 2}{\vec\sigma}\cdot\vec S+{\cal V}\right)a_{k\tau}^\dag 
a_{k'\tau}\hbox{\Huge{]}}.
\end{eqnarray}
In this equation, the operators $(\vec{S},\vec{\sigma})$ act on 
spin and the $(\vec{T},\vec{\tau})$ act on the (charge) orbital
degrees of freedom. 

The key role of this mapping stems from the
fact that $\langle\hat{Q}\rangle$ can be
identified as (an orbital pseudo-spin)
\begin{equation}\label{charge}
\langle\hat{Q}\rangle = e N_{grain} =
e \left({1\over 2} + \langle T^z \rangle\right).
\end{equation}
Then, we can introduce the extra (charge) 
state $|Q\rangle$ 
as an auxiliary label to the state $|\Phi\rangle$ of the grain. In addition to introducing the label $|Q\rangle$ we also make the replacement
\begin{eqnarray}
 a_{k1}^\dag a_{k'0}\hat P_0 &\longrightarrow&  a_{k1}^\dag a_{k'0}T^{+}
\\ \nonumber
a_{k0}^\dag a_{k'1}\hat P_1 &\longrightarrow& a_{k'0}^\dag a_{k1}T^{-},
\end{eqnarray}
following Ref. \onlinecite{Georg}.
Keep in mind that $T^{+}$ and $T^{-}$ are pseudo-spin ladder operators acting only
on the charge part $|Q\rangle$. We have the correct 
identifications 
\begin{eqnarray}
T^{-}|Q=1\rangle &=& T^{-}|T^z=+1/2\rangle\ = |Q=0\rangle \\ \nonumber
T^{+}|Q=0\rangle &=& T^{+}|T^z=-1/2\rangle\ = |Q=1\rangle,
\end{eqnarray}
meaning that the charge on the single-electron box is adjusted whenever a 
tunneling process takes place.
Furthermore, since $T^{+}|Q=1\rangle\ =0$ and  $T^{-}|Q=0\rangle\ 
=0$ these operators ensure 
in the same way as the projection operators $\hat P_0$ and 
$\hat P_1$ that only transitions between states with $Q=0$ and $Q=1$ take 
place. This leads us to identify $\hat{P}_1+\hat{P}_0$ with the identity 
operator on the
space spanned by $|0\rangle$ and $|1\rangle$ and $\hat{P}_1-\hat{P}_0$ with $2T^z$.
We now introduce an additional pseudo-spin operator
via:
\begin{eqnarray}
a_{k1}^\dag a_{k'0} &=& \frac{1}{2} 
a_{k\tau}^\dag\tau^- a_{k'\tau'}\\ \nonumber
a_{k0}^\dag a_{k'1} &=& \frac{1}{2} a_{k\tau}^\dag{\tau^+}a_{k'\tau'},
\end{eqnarray}
where the matrices $\tau^{\pm}=\tau^x\pm i\tau^y$ are standard combinations
of Pauli matrices. Bear in mind that here the operators $\hat{P}_{1,0}=(1\pm
2T^z)/2$ and $\hat{p}_{0,1}=(1\pm\tau^z)/2$ project out the grain state
with $Q=e$ and $Q=0$, and the reservoir/grain electron channels, 
respectively. 


When only ``charge flips'' are involved through the $V$ term, 
the model can be mapped onto a two-channel Kondo model (the two channels
correspond to the two spin states of an electron).\cite{Matv1} Here, we have a combination of spin and charge flips. A  question arises: can we expect two 
distinct energy scales for the spin and orbital sectors? To answer properly this question it is convenient
to rewrite the Kondo interaction in real space as
\begin{eqnarray}\label{heff}
\hskip -0.5cm H_K &=& {J\over 2}\vec{S}\cdot \left(\psi^{\dag}{\vec{\sigma}}\psi\right)+{V_{\perp}\over 2}\left[T^+ \left(\psi^{\dag}{\tau^-}\psi\right) +h.c.\right]\\ \nonumber
&+& Q_{\perp}\vec{S}\cdot \left[T^+(\psi^{\dag}{\tau^-}{\vec{\sigma}}\psi)
+h.c.\right],
\end{eqnarray}
where $\psi_{\tau\sigma}=\sum_k a_{k\tau\sigma}$ and the bare values are given by
\begin{equation}
\label{couplings}
V_{\perp}={\cal V}\ ,\  Q_{\perp}=J/4.
\end{equation}
Similar to a small dot coupled to two macroscopic leads, we have ignored the  potential scattering $V\psi^{\dag}\psi$ which does not renormalize and therefore will not affect the SU(4) theory at the
strong coupling fixed point.

\subsection{Enlarged symmetry in the infra-red}

We emphasize that a host of physical spin-exchange $\otimes$ isospin-exchange interactions 
are generated; $J$ refers to pure
spin-flip processes involving the S=1/2 spin of the small dot, 
$V_{\perp}$ to pure charge flips which modify the grain charge, and 
$Q_{\perp}$ describes exotic spin-flip assisted tunneling from leads to grain or vice-versa.
and those processes have been explicitly drawn in Ref. \onlinecite{Karyn}.

We note some discrepancies with Ref. \onlinecite{Schiller} concerning some bare values of the parameters as well as the form of the Renormalization Group (RG) equation flow.  In particular, our Hamiltonian exhibits a structure which is very similar to the one introduced in Ref. \onlinecite{Borda}
in order to study a symmetrical double (small) quantum dot structure 
with strong capacitive coupling. Two terms will be generated during the RG procedure, namely
\begin{equation}
{V_z\over 2}T^z \left(\psi^{\dag}
{\tau^z}\psi\right)\ ,\  Q_z T^z\vec{S}\cdot \left(\psi^{\dag}{\tau^z}{\vec{\sigma}}\psi\right),
\end{equation}
which fully respect the intrinsic symmetry of the bare Hamiltonian (A6).
We have implicitly assumed that $t\approx t_g$ such that $Q_z=0$ at the bare level. At the bare level we also get $V_z=0$. This bare condition differs from that used by the authors of Ref. \onlinecite{Schiller}. Moreover, focussing on their Hamiltonian (28), at the bare level we should have $\tilde{V}=0$ which hence should give $V_0=0$ in their Eq. (62).

For the particle-hole asymmetric case for the small dot where $V_{\perp}>0$ and large $U\gg -2\epsilon$, similar to Ref. \onlinecite{Borda} the RG flow\cite{Karyn} indicates that due to the presence of the spin-flip assisted tunneling terms $Q_{\perp}$ and $Q_z$ even though we start with very asymmetric bare values of the coupling constants all couplings diverge at the same energy scale
\begin{equation}
k_B T_K^{SU(4)}\sim D\ e^{-1/4\nu J},
\end{equation}
which is clearly enhanced due to the large underlying symmetry in the problem at low energy; $\nu$ is the density of states obtained after linearization of the electron dispersion relation.
Furthermore, a Numerical Renormalization Group (NRG) analysis has demonstrated that the SU(4) fixed point is the the appropriate fixed point for all allowed values of $V_{\perp}$ from $-J/4$ to $J/4$.\cite{Karyn} This reflects the
strong statement that when the spin-flip assisted tunneling terms $Q_{\perp}$ and $Q_z$, which strongly entangle the dot's spin $\vec{S}$ and the orbital spin of the grain $\vec{T}$, are relevant the only permissible fixed point by symmetry is the SU(4) model of Eq. (13) involving the hyper-spin operator
\begin{equation}
\left(\vec{S}+{1\over 2}\right)
 \left(\vec{\hbox{T}}+{1\over 2}\right).
\end{equation}

\subsection{Important remarks}

In short, when $t\approx t_g$, for a small dot in the Kondo regime implying $U\gg E_c$ and a grain at the degeneracy point, the SU(4) Kondo fixed point arises at low energy (both for a particle-hole symmetric and asymmetric level). Note that the NRG procedure used in Ref. \onlinecite{Schiller} could not explore the present limit where $U\gg E_c$. NRG clearly confirms that the SU(4) Fermi liquid fixed point resists for quite weak external magnetic field.\cite{Karyn} But, applying a {\it strong} magnetic field such that $g\mu_B B\gg k_B T_K^{SU(4)}$ unavoidably destroys the SU(4) symmetry. In a large magnetic field spin flips are suppressed at low temperatures, {\it i.e.}, $Q_{\perp}=Q_z=J=0$, and the orbital degrees of freedom,
through $V_{\perp}$ and $V_z$, can still develop a standard one-channel Kondo model
(the electrons have only spin-up or spin down), however the emerging Kondo temperature 
will be much smaller, 
\begin{equation}
k_B T_K[B=\infty]\approx D \ e^{-1/{\cal V}},
\end{equation}
with for instance ${\cal V}\approx t t_g/(-2\epsilon)$ for $U\rightarrow +\infty$,
and might not be detectable experimentally. A substantial
decrease of the Kondo temperature when applying an
external magnetic field $B$ has also been certified using NRG.\cite{Karyn}

\section{Conformal field theory}\label{sec:cft}

\subsection{The strong coupling fixed point}
The bulk theory may be described by $4$ free spinless chiral fermions $\psi_j$ 
embodied by the Hamiltonian
density
\begin{equation}
{\cal H}_0=i\frac{v_F}{2\pi}\sum\limits_{j=1}^4 \psi^\dagger_j \frac{d}{dx} \psi_j.
\end{equation}
In this Appendix, $\psi_j$ must be identified as $\psi_j(x)$ and $\psi_j$ should be understood as the left moving fermion when the index L is omitted ($v_F$ is the Fermi velocity). Note the unconventional normalization similar to that of Ref. \onlinecite{AL}. Along the lines of Affleck and Ludwig\cite{AL} we  introduce the charge and hyperspin operators
\begin{eqnarray}
J_c= \psi^{j\dagger}\psi_j \\ \nonumber
J^A = \psi^{i\dagger}(T^A)_i^j \psi_j.
\end{eqnarray}
Here, $T^A$ are a Cartan basis of generators of $SU(4)$
and the sum over the indices follows the Einstein convention.
From conformal theory arguments, we can rewrite the $4$ fermion theory in terms of the {\it charge}
and {\it hyper-spin} operators. The central charge of the $SU(4)_{k=1}$ group is
\begin{equation}
c=\frac{dim(SU(4)).1}{4+1}=15/5=3.
\end{equation}
Provided we add the charge sector with central charge $1$, then $c=3+1=4$, as it should be
for $4$ free fermions. 

To know the exact correspondence one can proceed as follows.  A straighforward calculation results in
\begin{equation}
J_c^2= : \psi^{i\dagger}\psi_i\psi^{j\dagger}\psi_j : +2i \psi^{j\dagger} \frac{d}{dx} \psi_j.
\end{equation}
Moreover,
\begin{equation}
J^A J^A = \sum_A \psi^{i\dagger} (T^A)_i^j \psi_j  \psi^{k\dagger} (T^A)_k^l\psi_l.
\end{equation}
Now, one can resort to
\begin{equation}
\sum_A (T^A)_i^j (T^A)_k^l = \frac{1}{2}(\delta_k^j \delta_i^l - \frac{1}{4}\delta_i^j\delta_k^l). 
\end{equation}
One immediately infers:
\begin{equation}
J^A J^A = -\frac{5}{8} :\psi^{i\dagger}\psi_i\psi^{j\dagger}\psi_j : +\frac{15}{4}i \psi^{j\dagger} \frac{d}{dx} \psi_j.
\end{equation}
This leads to
\begin{equation}
{\cal H}_0= \frac{v_F}{2\pi}\left(\frac{1}{5}J^A J^A + \frac{1}{8}J_c J_c\right).
\end{equation}
Taking into account  the Kondo interaction $H_K$ defined in Eq. \ref{irrrep} the total Hamiltonian density 
${\cal H}$  takes the form
\begin{eqnarray}
{\cal H} &=& \left(\frac{v_F}{16\pi}J_c J_c\right) \\ \nonumber
&+& \left(\frac{v_F}{10\pi}J^A J^A+\frac{J_K}{4} J^A M^A\delta(x)\right);
\end{eqnarray}
$M^A$ is the localized hyper-spin. As a reminiscence of the SU(2) case, one can complete the square at the value:
\begin{equation}
J_{K}^* = \frac{4v_F}{5\pi}.
\end{equation}
In the strong coupling realm, implying $T\ll T_K^{SU(4)}$, the SU(4) impurity may be totally absorbed in ${\cal H}_0$ with new boundary conditions on the fields $\psi_j$.  This is equivalent to say that at the strong coupling fixed point $J_{K}^*$ the hyper-spin $M^A$ is completely screened. 

Now, let us analyze  the vicinity of the strong coupling fixed point, thoroughly.

\subsection{Spectroscopy at the strong coupling fixed point}\label{sec:spectro}

 Similar to the SU(2) case, the leading irrelevant operator allowed at the strong coupling fixed point is 
the dimension-2 rotationally invariant operator  
\begin{equation}
\lambda J^A(0) J^A(0) = \lambda J^A J^A \delta(x),
\end{equation}
which we write as
\begin{eqnarray}
\lambda J^A(0) J^A(0) = -\frac{5}{8} \lambda  :\psi^{i\dagger}(0)\psi_i(0)\psi^{j\dagger}(0)\psi_j(0) : \\ \nonumber
+\frac{15}{4}\lambda i \psi^{j\dagger}(0) \frac{d}{dx} \psi_j(0),
\end{eqnarray}
and $\lambda$ is of the order of $1/k_BT_K^{SU(4)}$.

{\it Let us first focus  on the elastic part.} One can write the change
in the boundary condition as\cite{AL}
\begin{equation}
<\psi^\dagger_{Lj}(z_1) \psi_{Rj}(\bar{z}_2)>\ = \frac{e^{2i\delta}}{z_1-\bar{z}_2}=\frac{S(1)}{z_1-\bar{z}_2},
\end{equation}
where we have restored the chiral indices and $S(1)$ is a universal complex number which depends only on the universality class of the boundary conditions. This represents the scattering
matrix. In the SU(4) case,
\begin{equation}
S(1) = e^{2i\delta} = e^{i\pi/2}.
\end{equation}
We use the standard conventions $\psi_{Lj}(x,\tau)=\psi_{Lj}(z=\tau+ix)$ and $\psi_{Rj}(x,\tau)=\psi_{Rj}(\bar{z}=\tau-ix)$, that means
\begin{equation}
\psi_{Rj}(\bar{z})=S(1)\psi_{Lj}(z)=i\psi_{Lj}(z).
\end{equation}

By analogy with Affleck and Ludwig,\cite{AL} we can derive the first elastic correction to the
retarded self-energy $\Sigma^R$:
\begin{equation}
\Sigma^R(\omega) = -\frac{1}{2\pi \nu}[i(1-e^{i\pi/2})+i\frac{15}{2}\pi \lambda\omega],
\end{equation}
as opposed to the SU(2) case where
$
\Sigma_{SU(2)}^R(\omega) = -\frac{1}{2\pi \nu}[2i-3\pi \lambda\omega]$. The first term stems from
\begin{equation}
\Im m\Sigma^R(\omega) = -\frac{i}{2\pi \nu}[1-S(1)].
\end{equation}
One can find the effective phase-shift by rewriting
\begin{equation}
\Sigma^R(\omega) = -\frac{i}{2\pi \nu}[1-e^{2i\delta(\omega)}].
\end{equation}

In the SU(4) case, this  gives
\begin{eqnarray}
\Sigma^R(\omega) &=& -\frac{i}{2\pi \nu}[1-e^{i\pi/2}(1+i\frac{15}{2}\pi \lambda\omega)] \\ \nonumber
&=& -\frac{i}{2\pi \nu}[1-e^{i\pi/2}e^{i\frac{15}{2}\pi\lambda\omega}].
\end{eqnarray}
We therefore infer
\begin{equation}
\delta(\omega)=\frac{\pi}{4} +\frac{15}{4}\pi\lambda\omega,
\end{equation}
which should be explicitly contrasted with the well-known SU(2) result $\delta_{SU(2)}(\omega)=\frac{\pi}{2} +\frac{3\pi\lambda}{2}\omega$.

Following Affleck and Ludwig,\cite{AL} one may compute the first  {\it inelastic} contribution
to the self-energy which is of ${\cal O}(\lambda^2)$.
The lowest order inelastic contribution reads:
\beq
\Sigma^R_{inel}=\frac{-1}{2\pi\nu}\left(\frac{5\pi\lambda}{4}\right)^2(\omega^2+\pi^2T^2).
\eeq
The fact that $\Sigma^R_{inel}$ is real to second order in $\lambda$ stems from
the additional multiplicative factor $e^{2i\delta}=i$ in the case of the SU(4) Kondo correlated fixed point.
Gathering the elastic part to ${\cal O}(\lambda^2)$ and the inelastic part, we thus obtain:
\bea
\Sigma^R(\omega)&=& -\frac{i}{2\pi \nu}\left[1-e^{i\pi/2}+\frac{15}{2}\pi \lambda\omega\right.\\
&+&i\left.\left(\frac{5\pi\lambda}{4}\right)^2(17\omega^2-\pi^2T^2)+{\cal O}(\lambda^3)\right].\nn
\eea

Let us focus on $-\hbox{Im}{\cal T}_{\sigma}$ or
equivalently $-\hbox{Im}\Sigma^R(\omega)$. The contribution to the T-matrix is simply
\begin{eqnarray}
-\pi \nu \hbox{Im}{\cal T}_{\sigma} &=& \frac{1}{2}\hbox{Re}\left[1-e^{i\frac{\pi}{2}+\frac{15}{2}\pi\lambda\omega}\right] \\ \nonumber
&=& \frac{1}{2}\left(1+\frac{15}{2}\pi\lambda\omega\right)+{\cal O}(\lambda^3).
\end{eqnarray}
\vskip 0.3cm
One thus infer
\begin{equation}
G(V,T) \approx G_0\left(\frac{1}{2} +\frac{15}{4}\pi\lambda eV+{\cal O}(\lambda^3)\right),
\end{equation}
where $\lam\sim 1/k_B T_K^{SU(4)}$.
This result implies in particular that the temperature corrections to the zero bias conductance
are of the order of $T^3$, {\it i.e.}, much weaker than in the SU(2) Kondo Fermi liquid realm, which implies
\beq
G(V,T)-G(V,0) \approx a \left(T/T_K^{SU(4)}\right)^3,\eeq
 where $a$ is a (negative) constant. 
 In order to determine the constant parameter $a$ one needs to compute the third order diagrams that are quite tedious to evaluate.


\subsection{Asymmetry between channels}\label{sec:asymmetry}

A channel asymmetry can be summarized as:
\begin{equation}
\delta {\cal H} = {\delta J}
 \sum_{i,j} \psi^{\dagger}_i \left(\vec{\sigma}\cdot \vec{S}\right) \tau^z\psi_j \delta(x).
\end{equation}
It is then convenient to resort to the orbital-spin decomposition $i=(\tau,\sigma)$ of Appendix A, with $(\tau=0,1)$ and $\sigma=(\uparrow,\downarrow)$. It is important to observe that
 to {\it first} order in $\delta J$ the main contribution takes the form
\begin{equation}
\delta {\cal H} \sim \delta J \sum_{\tau} (\psi^{\dagger}_{\tau\uparrow}\psi_{\tau\uparrow}-\psi^{\dagger}_{\tau\downarrow}\psi_{\tau\downarrow})S^z\tau^z \delta(x).
\end{equation}
In particular, this will provide an extra phase-shift to the $a_{k\sigma}$ electrons. More precisely, at the strong coupling fixed point we can use the precise
identification
\begin{equation} 
(\psi^{\dagger}_{\tau\uparrow}\psi_{\tau\uparrow}-\psi^{\dagger}_{\tau\downarrow}\psi_{\tau\downarrow})S^z
\approx -\frac{1}{2} (\psi^{\dagger}_{\tau\uparrow}\psi_{\tau\uparrow}+\psi^{\dagger}_{\tau\downarrow}\psi_{\tau\downarrow}).
\end{equation}
Thus, this results in
\begin{eqnarray}
\delta {\cal H} &=&-\frac{\delta J}{2}\sum_{\tau} \tau^z (\psi^{\dagger}_{\tau\uparrow}\psi_{\tau\uparrow}+\psi^{\dagger}_{\tau\downarrow}\psi_{\tau\downarrow}) \delta(x)\\ \nonumber
&=& -\delta J \sum_i \psi^\dagger_i \frac{\tau^z}{2}\psi_i \delta(x).
\end{eqnarray}
To the lowest (first) order in $\delta J$, the channel asymmetry is thus equivalent to an effective (local) magnetic field in the orbital sector. This observation has been precisely confirmed through NRG.\cite{Karyn}


\begin{thebibliography}{99}
\bibitem{note1}
In particular, for $|\delta J|\gg 0$, the orbital (2-channel) Kondo effect between the grain and the small dot would occur at unrealistically low temperatures.\cite{Karyn}
\end{thebibliography}
\end{document}